\documentstyle[12pt]{article}
\newcommand{\be}{\begin{equation}}
\newcommand{\nn}{\nonumber}
\newcommand{\bea}{\begin{eqnarray}}
\newcommand{\eea}{\end{eqnarray}}
\newcommand{\ba}{\begin{array}}
\newcommand{\ea}{\end{array}}
\newcommand{\ee}{\end{equation}}

\expandafter\ifx\csname mathbbm\endcsname\relax

\else

\fi \textheight 22cm \textwidth 15cm \topmargin 1mm \oddsidemargin
5mm \evensidemargin 5mm

\newcommand{\beas}{\begin{eqnarray*}}
\newcommand{\eeas}{\end{eqnarray*}}
\newcommand{\bes}{\begin{equation*}}
\newcommand{\ees}{\end{equation*}}

\newcommand{\lf}{\left}
\newcommand{\ri}{\right}
\newcommand{\f}{\frac}

\def\tr           {\mbox{\rm tr}\,}

\def\i2           {\mbox{$\frac{i}{2}$}}

\def\del           {\delta}

\def\ep           {\epsilon}

\def\la           {\lambda}

\def\si           {\sigma}

\begin{document}

\begin{titlepage}
\hfill \vbox{
    \halign{#\hfil         \cr
           } 
      }  
\vspace*{20mm}
\begin{center}
{\LARGE \bf{{Quasi-Instantons and Supersymmetry Breaking}}}\\ 

\vspace*{15mm} \vspace*{1mm} {Ali Imaanpur}

\vspace*{1cm}

{\it Department of Physics, School of Sciences \\
Tarbiat Modares University, P.O. Box 14155-4838, Tehran, Iran\\
Email: aimaanpu@theory.ipm.ac.ir}\\
\vspace*{1mm}

\vspace*{1cm}

\end{center}

\begin{abstract}
We show that supersymmetric gauge theories with a product gauge group admit quasi-instantons, 
which, like instantons, sit at the absolute minimum of the action in the corresponding topological 
sector. However, unlike ordinary instantons which preserve part of the supersymmetry, these  
break all supersymmetries of the action. We also provide a generalization where quasi-instantons   
take value in the associated Lie algebra of the recently proposed 3-algebras. 
When the structure constants and the metric of the algebra are covariantly constant, the gauge bundle 
admits new topological sectors with quasi-instantons at the minimum of the action.  
\end{abstract}

\end{titlepage}

\section{Introduction}
Gauge theories with simple unitary gauge groups fall into distinct topological sectors characterized 
by an integer instanton number over which one has to sum up to get 
the partition function of the theory in Euclidean space. However, when the gauge group 
is a direct product of two simple unitary gauge groups,\footnote{The discussion can easily be 
extended to gauge groups of more than two simple factors.}  the gauge bundle 
is characterized by two such instanton numbers; one for each gauge group factor. In other words, 
since there are two ways of invariant contracting of the gauge algebra indices, we can write 
down two topological invariants for the gauge bundle. For each such an invariant we can obtain a 
lower bound on the action which is saturated either for (anti)instantons or (anti)quasi-instantons. 

For gauge theories associated with a 3-algebra \cite{BAG1}, a similar situation in four-dimension 
arises. Given a 3-algebra, one can write down the associated Lie algebra, on which the gauge 
fields take value, i.e., $A=A_{ab}T^{ab}$, with $T^{ab}$ the generators. Interestingly, with the use 
of a fundamental identity, one can show that the structure 
constants, $f^{abc}_{\ \ \ d}$, of the 3-algebra are invariants of the associated Lie algebra. The algebra 
can also be endowed with an invariant positive definite metric $h^{ab}$. Having been equipped with these two 
invariants of the Lie algebra, we can use them to contract the Lie algebra indices. Further, if we 
require the structure constants and metric be covariantly constant, we can write a new topological 
term: 
\be
\int f^{abcd}\ F_{ab}\wedge F_{cd}\, ,
\ee
which, in turn, indicates the appearance of new topological sectors in the gauge theory. 
We will focus on cases where this term sets a lower bound on the action with the minimum value 
attained by quasi-instantons:
\be
F^{ab}=\f{1}{2}\, f^{abcd}\, * F_{cd}\, . 
\ee
The admixture of the Lie algebra indices through $f^{abcd}$ on the right hand side has an important 
implication for supersymmetry. For the special case of $f^{abcd}\sim \varepsilon^{abcd}$, where 
the associated Lie algebra is $so(4)\approx su(2) \oplus su(2)$, we will 
see that these configurations, in contrast with instantons, break all supersymmetries of the action.  

In the next section, we discuss the possibility of having quasi-instantons when the gauge group is a 
direct product of two unitary groups. We will see that they sit at 
the minimum of the action in topological sectors where two gauge factors have instanton numbers of  
opposite signs. In section 3, we propose a generalization of quasi-instantons using the structure constants 
of 3-algebras. We write a topological term similar to the Euler class, and argue that this term sets a 
lower bound on the action of quasi-instantons. In section 4, we move on to the discussion of supersymmetry. 
It is observed that quasi-instantons classically break all supersymmetries of the action. Since these 
configurations are at the absolute minimum of the action it is quite plausible that supersymmetry is 
spontaneously broken. In this case, it is important to investigate the implications for phenomenological 
theories like supersymmetric quiver gauge theories.

\section{Product Gauge Groups and Quasi-Instantons}
Let us start our discussion with the following argument which is usually used to show that instantons 
sit at the absolute minimum of the action and thus are solutions to the equations of motions. 
First notice that
\be
\int \tr\lf[(F\pm * F)\wedge *(F\pm * F)\ri]\, \geq 0 \, , 
\ee
where $*$ indicates the Hodge star. This implies
\be
\int \tr(F\wedge * F)\, \geq\, \mp \int \tr(F\wedge F)=\mp k\, ,
\ee
for $k$ the instanton number.\footnote{We drop a factor of $\f{1}{16\pi^2}$ from the definition of $k$ for 
clarity in the equations.} The bound is saturated for (anti)instanton configurations
\be
F^\pm=\f{1}{2}\, (F\pm * F)=0\, .\label{INST}
\ee
Notice that for (anti)instantons $k$ is a (positive) negative integer, and hence 
the minimum of Yang-Mills action is $|k|$ for either instantons or anti-instantons.

To write down the quasi-instanton equations, we need to use the explicit algebra 
indices. Let $J^A$ indicate the generators of the gauge group algebra. The gauge field and the field 
strength take value in the Lie algebra
\be
A=A^AJ^A\, ,\ \ \ \ F=F^AJ^A\, .
\ee
We use a normalization of the generators such that
\be
\tr(J^AJ^B)=\del^{AB}\, ,
\ee
where $A=(a,b)$ is a collective index; $a$ and $b$ take value in the two subalgebras.
For $k$ we write
\be
k= \int \tr(F\wedge F)= \int F^A\wedge F^A \label{TOP}
\ee
and the instanton equations (\ref{INST}) read
\be
F^A_{\mu\nu}=\mp\f{1}{2}\ep_{\mu\nu\rho\si}F_{\rho\si}^A \label{INS}\, .
\ee
Writing this way shows that the bound is saturated when we have {\em only} instantons (anti-instantons) 
for {\em both} gauge groups. 

For gauge theories for which the gauge group is a direct product of two simple gauge groups, 
apart from (\ref{TOP}), there is yet another topological invariant. In fact, since the Lie 
algebra is the direct sum of two simple Lie algebras, one can write down the following invariant 
trace form 
\be
k'= \int { {\tr}'}(F\wedge F)= \int F^a\wedge F^a -\int F^b\wedge F^b \, ,\label{MIN}
\ee
where we have split the Lie algebra index $A=(a,b)$. Let us now repeat our argument 
from the beginning of this section. The topological term in (\ref{MIN}) allows us to get a bound on 
the action
\be
\int { {\tr}}(F\wedge *F)= \int F^a\wedge *F^a +\int F^b\wedge *F^b \, .
\ee
Note that
\be
\int \lf[(F^a\pm * F^a)\wedge *(F^a\pm * F^a)\ri]+\lf[(F^b\mp * F^b)\wedge *(F^b\mp * F^b)\ri]\, \geq 0 \, , 
\ee
where we have swapped the signs in the second term to account for the minus sign in (\ref{MIN}). 
Therefore
\be
\int F^a\wedge *F^a +\int F^b\wedge *F^b \geq \mp \int F^a\wedge F^a \pm\int F^b\wedge F^b
\ee
so the bound is saturated if either
\be
F^a_{\mu\nu}=-\f{1}{2}\ep_{\mu\nu\rho\si}F_{\rho\si}^a \, , \ \ \ \ F^b_{\mu\nu}=+\f{1}{2}\ep_{\mu\nu\rho\si}F_{\rho\si}^b \, ,
\label{Q1}
\ee
or
\be
F^a_{\mu\nu}=+\f{1}{2}\ep_{\mu\nu\rho\si}F_{\rho\si}^a \, , \ \ \ \ F^b_{\mu\nu}=-\f{1}{2}\ep_{\mu\nu\rho\si}F_{\rho\si}^b \, .
\label{Q2}
\ee
We call these quasi-instantons, as in contrast with instantons (\ref{INS}), here we have instantons for 
one gauge subalgebra, and anti-instantons for the other. It is also obvious that quasi-instantons, 
(\ref{Q1}) and (\ref{Q2}), satisfy the equation of motion, $D_\mu F^A_{\mu\nu}=0$. The action is 
minimized for quasi-instantons
\be
S= |k_1|+|k_2|\, ,
\ee
where
\be
k_1=\int F^a\wedge F^a \, ,\ \ \ \ k_2=\int F^b\wedge F^b \, .
\ee

\section{Generalization via 3-algebras}

Triple algebras have naturally emerged in searching for a gauge invariant action describing 
multiple M2-branes \cite{BAG1, GUS}. In this construction the only free parameters in the action are 
the structure constants of the 3-algebra. However, it has been shown that 3-algebras with 
totally antisymmetric structure constants are reducible to the direct sum of a unique 3-algebra 
with the structure constants proportional to $\varepsilon_{abcd}$ \cite{PAPA, GAUNT}. 
To get more general 3-algebras, it is therefore necessary to relax some constraints on the structure constants 
(which consequently results in different actions of M2-branes) \cite{BAG2}. In the following, first 
we begin with a discussion of the 3-algebras with totally antisymmetric structure constants, and then 
consider the more general case. We will see that the 3-algebras with a covariantly constant structure 
constant and metric allow us to write down a new topological term of the gauge bundle, indicating 
the appearance of new topological sectors in the gauge theory. Further, we can  
generalize the quasi-instanton equations of section 2. 

Let us begin with a brief review of 3-algebra definition, we refer to \cite{BAG1} for details. Let $T^a$'s 
indicate the generators of the algebra with $a=1, \ldots , N$.\footnote{ The $a$ and $b$ indices in this section 
refer to the 3-algebra indices and should not be confused with the Lie algebra indices of the previous 
(and also the next) section.}  The triple algebra 
is defined by 
\be
[T^a,T^b,T^c]=f^{abc}_{\ \ \ d}\, T^d \, ,
\ee
where the structure constants $f^{abc}_{\ \ \ d}$ satisfy the fundamental identity
\be
f^{efg}_{\ \ \ d}f^{abc}_{\ \ \ g}= f^{efa}_{\ \ \ g}f^{bcg}_{\ \ \ d}+ f^{efb}_{\ \ \ g}f^{cag}_{\ \ \ d}+ 
f^{efc}_{\ \ \ g}f^{abg}_{\ \ \ d} \, .\label{FU}
\ee
The indices can be raised (lowered) with the following positive-definite metric
\be
h^{ab}=\tr (T^a,T^b)\, ,
\ee
then it is shown that $f^{abcd}=f^{abc}_{\ \ \ e}h^{ed}$ is totally antisymmetric.

Following \cite{BAG1}, the one-form gauge field is defined through 
\be
A^a_{\ b}=f^{cda}_{\ \ \ b}A_{cd}\equiv (T^{cd})^a_{\ b}A_{cd}\, ,
\ee
where we have introduced the matrices  
\be
(T^{cd})^a_{\ b}= f^{cda}_{\ \ \ b} 
\ee
to identify the algebra on which the gauge fields take value. 
Now using the fundamental identity (\ref{FU}), it is easy to see that $T^a$'s form a closed Lie algebra:
\be
[T^{ab},T^{cd}]= -f^{cda}_{\ \ \ e}T^{be}+ f^{cdb}_{\ \ \ e}T^{ae}
+f^{abc}_{\ \ \ e}T^{de}- f^{abd}_{\ \ \ e}T^{ce}\, ,\label{G}
\ee 
where we have antisymmetrized with respect to $ab \leftrightarrow cd$ and then rescaled the generators.\footnote
{A similar algebra with a central charge extension has also appeared in an infinite dimensional generalization 
of 3-algebras \cite{LIN}.} 
Note that the fundamental identity (\ref{FU}) implies that $f^{abcd}$ is an invariant 4-form of the above 
Lie algebra. 

Dropping the matrix indices, we write the one-form gauge fields as
\be
A = A_{ab}T^{ab}\, ,
\ee
so that the field strength reads
\be
F= dA+ A\wedge A= F_{ab}T^{ab}\, ,
\ee
which satisfies the Bianchi identity:
\be
DF=dF+[A,F]=0\, .
\ee

Since $f^{abcd}$ are invariant 4-form of algebra (\ref{G}), in writing an action for the gauge fields, 
there are two ways of invariant contracting of the algebra indices. In particular, we can form two 
topological invariants of the gauge bundle. The first one is the ordinary second Chern class, and 
the corresponding instanton number (using the metric and a proper normalization of the generators):
\be
\int F^{ab}\wedge F_{ab}\, ,
\ee
sets a lower bound for ordinary instantons
\be
F^{ab}=*\, F^{ab}\, .
\ee
As for the second topological term, we write: 
\be
\int f^{abcd}\, F_{ab}\wedge F_{cd}\, ,\label{TOPP}
\ee
which is similar to the Euler characteristic of the $SO(N)$ frame bundle.

Now, to write the quasi-instanton equations, we need to put further constraints on the structure constants. 
Consider the matrix $f^{abef}f^{cd}_{\ \ ef}$ which is symmetric in $ab\leftrightarrow cd$. 
Suppose this matrix has no degenerate eigenvalues and thus can be diagonalized. Further let us 
assume that it is non-degenerate. The structure constants and the generators can then be rescaled 
such that we have
\be
f^{abef}f_{cdef}= 2\, (\del^{a}_{\ c}\del^{b}_{\ d}-\del^{a}_{\ d}\del^{b}_{\ c})\, .\label{REL}
\ee
This relation allows us to define a Hodge star like operation, $\tilde{*}$, on the Lie algebra indices
\be
(\tilde{*}F)^{ab}=\f{1}{2}\, f^{abcd}\, F_{cd}\, ,
\ee
which, through using (\ref{REL}), squares to one
\be
\tilde{*}^2 F= F\, .
\ee
Hence we can write the quasi-instanton equations as follows:
\be
F^{ab}=\pm \f{1}{2}\, f^{abcd}\, * F_{cd}\, ,\label{QUAS}
\ee
or more succinctly as
\be
\tilde{*}\, F=\pm * F\, .
\ee
Note that for these to solve the equations of motion, the structure constants $f^{abc}_{\ \ \ d}$ 
and the metric $h^{ab}$ on the triple algebra are needed to be covariantly constant. This condition is 
equivalent to saying that the term (\ref{TOPP}) is a genuine topological term. Also notice that, in 
writing the topological term, (\ref{TOPP}), or the equations, (\ref{QUAS}), the structure 
constants are not needed to be totally antisymmetric, we only require
\be
f^{abcd}=-f^{bacd}\, , \  \  f^{abcd}=-f^{abdc}\, , \ \  f^{abcd}=f^{cdab}\, \  \, ,
\ee 
which are the same constraints appearing in the generalized 3-algebras \cite{BAG2, CHE}.

The squaring argument of the previous section shows that the quasi-instantons, (\ref{QUAS}), have a 
minimum action of
\be
\int F^{ab}\wedge * F_{ab} = \mp \int f^{abcd}\, F_{ab}\wedge F_{cd}\, .
\ee
For the special case of $f^{abcd}\sim \varepsilon^{abcd}$, the associated Lie algebra is 
$so(4)\approx su(2)\oplus su(2)$, and therefore eqs. (\ref{QUAS}), with an appropriate change of basis, 
will reduce to eqs. (\ref{Q1}) and (\ref{Q2}) of section 2.

\section{Supersymmetry Breaking via Quasi-Instantons}
In this section we will examine the supersymmetry breaking in theories with a product gauge group,  
and with some matter field in the bifundamental representation (like quiver gauge theories). 
In the absence of matter field, the supersymmetry of the action can be enhanced to two supersymmetries,  
one for each gauge factor, and the following argument for supersymmetry breaking can be 
evaded.

It is well known that in supersymmetric gauge theories on Euclidean space instantons preserve 
half of supersymmetries. Explicitly, a non-trivial supersymmetric background can be   
obtained by setting fermions to zero and then demanding supersymmetry variation of fermions vanish. 
Consider, for instance, $N=1$ supersymmetry transformations of fermions. Setting $D$-terms to zero we have:
\be
\del \la^A= \si^{\mu\nu}\ep\, F^A_{\mu\nu} \, ,\label{S1}
\ee
and for ${\bar \la}$
\be
{\del} {\bar \la}^A= {\bar\ep} {\bar \si}^{\mu\nu}F^A_{\mu\nu} \, .\label{S2}
\ee  
As $\si_{\mu\nu}$ (${\bar \si}_{\mu\nu}$) are (anti)self-dual, and since on Euclidean space 
$\ep$ and ${\bar \ep}$ are independent of each other, we see that (anti)instantons are 
preserving half the supersymmetries. 

Now, let us consider quasi-instantons in (\ref{Q1}):
\be
F^{a+}_{\mu\nu}=0\, ,\ \ \ \ F^{b-}_{\mu\nu}=0\, ,
\ee
where we recall that $a$ and $b$ refer to the two subalgebra indices.
The supersymmetry transformations, (\ref{S1}) and (\ref{S2}), in this background read
\bea
&& \del \la^a=0 \nn \\
&& \del \la^b=  \si^{\mu\nu}\ep\, F^{b+}_{\mu\nu} \nn \\
&& {\del} {\bar \la}^a= {\bar\ep} {\bar \si}^{\mu\nu}F^{a-}_{\mu\nu}\nn \\ 
&& {\del} {\bar \la^b}=0 \, .
\eea  
Therefore it is seen that there is no way to preserve supersymmetry in this background; 
$Q$ and $\bar{Q}$ are both broken. Note that in computing the partition function or the correlation 
functions of the theory, we have to sum over all different 
topological sectors of the gauge bundle, and in so doing, we need to distinguish 
between two kinds of bundles depending on the sign of instanton numbers of the corresponding gauge 
groups: Bundles which are characterized by instanton numbers of the same sign, and those with 
opposite sign instanton numbers. In the former case, we do have classical supersymmetric backgrounds 
of (anti)instantons. However, in the latter case, we cannot have classical supersymmetric backgrounds. 
By turning on fermions and other scalars in the theory, it is unlikely to obtain supersymmetric 
solutions having the same action as that of quasi-instantons. Further, since quasi-instantons are 
{\em exact} classical solutions which break {\em all} supersymmetries, it seems a semi-classical computation 
of the path integral around these vacua (like the ones done in the case of instantons \cite{NOV, AMATI, VAN}) 
will result in a spontaneous supersymmetry breaking. It is therefore important to examine 
this mechanism of supersymmetry breaking in  phenomenological theories, like supersymmetric quiver gauge theories, 
which have the same kind of product gauge groups as the ones discussed above.

\newpage

\hspace{30mm}


\end{document}